\documentclass[sigconf]{acmart}

\usepackage{soul}
\usepackage{xcolor}
\usepackage{subcaption}
\usepackage{amsmath}
\usepackage{booktabs}
\usepackage{csquotes}

\newcommand{\subr}[1]{{\small\texttt{r/#1}}}
\newcommand{\madval}[1]{\mbox{\textsc{mad}\,=\,#1}}
\newcommand{\medval}[1]{\mbox{$\tilde{x}$\,=\,#1}}
\newcommand{\medmadval}[2]{\medval{#1}; \madval{#2}}
\newcommand{\Fhalf}{$F_{\text{½}}$}
\newcommand{\Fhalfval}[1]{\mbox{\Fhalf\,=\,{#1}}}

\copyrightyear{2023}
\acmYear{2023}
\setcopyright{acmlicensed}\acmConference[WebSci '23]{15th ACM Web Science Conference 2023}{April 30-May 1, 2023}{Austin, TX, USA}
\acmBooktitle{15th ACM Web Science Conference 2023 (WebSci '23), April 30-May 1, 2023, Austin, TX, USA}
\acmPrice{15.00}
\acmDOI{10.1145/3578503.3583626}
\acmISBN{979-8-4007-0089-7/23/04}

\begin{document}

\title[One of Many: Assessing User-level Effects of Moderation Interventions on r/The\_Donald]{One of Many: Assessing User-level Effects\\of Moderation Interventions on r/The\_Donald}

% COMMENTED FOR DOUBLE BLIND SUBMISSION
%\begin{comment}
\author{Amaury Trujillo}
\email{amaury.trujillo@iit.cnr.it}
\orcid{0000-0001-6227-0944}
\affiliation{%
  \institution{IIT-CNR}
  \country{Italy}
}

\author{Stefano Cresci}
\email{stefano.cresci@iit.cnr.it}
\orcid{0000-0003-0170-2445}
\affiliation{%
  \institution{IIT-CNR}
  \country{Italy}
}

\renewcommand{\shortauthors}{Trujillo \& Cresci}
%\end{comment}

%\author{Anonymous Author(s)}

\begin{abstract}
Evaluating the effects of moderation interventions is a task of paramount importance, as it allows assessing the success of content moderation processes. So far, intervention effects have been almost solely evaluated at the aggregated platform or community levels. Here, we carry out a multidimensional evaluation of the \textit{user-level} effects of the sequence of moderation interventions that targeted {\small\texttt{r/The\_Donald}}: a community of Donald Trump adherents on Reddit. We demonstrate that the interventions: \textit{1)} strongly reduced user activity; \textit{2)} slightly increased the diversity of the subreddits in which users participated; \textit{3)} slightly reduced user toxicity; and \textit{4)} gave way to the sharing of less factual and more politically biased news. Importantly, we also find that interventions having strong community level effects are associated to \textit{extreme and diversified user-level reactions}. Our results highlight that community-level effects are not always representative of the underlying behavior of individuals or smaller user groups. We conclude by discussing the practical and ethical implications of our results. Overall, our findings can inform the development of targeted moderation interventions and provide useful guidance for policing online platforms.
\end{abstract}





\maketitle

\section{Introduction}
\label{sec:intro}
For a few years now, online platforms have been facing public and governmental pressure to take action against online harms such as the spread of mis- and disinformation, and the rise of toxic and hateful speech. Platforms have responded to the growing pressure by deploying a multitude of moderation interventions, specific actions through which they intend to mitigate user misbehavior~\cite{gillespie2018custodians}. % performing content moderation and, specifically, by
As recent examples, Reddit, Facebook, Instagram, and Twitter attached warning labels to posts presenting disputed information about COVID-19 and election results~\cite{zannettou2021won,krishnan2021research}, and banned toxic users and communities~\cite{horta2021platform,jhaver2021evaluating,trujillo2022make}. On the one hand, these moderation interventions appear as reasonable solutions, representing initial countermeasures to many online harms, and serve as public evidence of the platforms’ willingness to tackle the issues they contributed to create. On the other hand, many of such interventions are applied light-mindedly and without proper validation~\cite{blaya2019cyberhate}. Content moderation is still mostly expert-driven, the design of interventions is based on ``common sense and intuition'', and progress is sought via trial-and-error rather than a rigorous scientific approach~\cite{cresci2022personalized}. Consequently, the effectiveness of current content moderation strategies is largely open to question and the need for additional evaluation efforts is manifest~\cite{singhal2022sok}. % research and experimentation

Evaluating the effects of moderation interventions is challenging. First, effects are \textit{multidimensional}: they affect multiple facets of user behavior, such as news consumption, interests, community participation habits, and polarization, to name a few~\cite{trujillo2022make}. However, existing research has almost solely considered content \textit{activity} and \textit{toxicity} when evaluating intervention effects~\cite{chandrasekharan2017you,chandrasekharan2022quarantined,jhaver2021evaluating,saleem2018aftermath,horta2021platform}. Plenty of other dimensions are essentially unexplored, meaning that we currently only have a partial view of the full extent of the effects caused by moderation interventions. Second, intervention effects can be evaluated at \textit{different levels}: from multiple or single platforms and communities, down to individual users. So far, the vast majority of works that evaluated intervention effects did so at the platform or community levels. However, aggregated effects at these levels are the combination of many and potentially diverse effects at the user level. Hence, such aggregated effects might not be truly representative of the underlying behavior of individuals or smaller user groups~\cite{welles2014minorities,robertson2022uncommon}. Moreover, the same aggregated effect could be the result of heterogeneous distributions of user-level effects~\cite{cresci2019semantically,gelman2023causal}, each corresponding to a different practical situation. Knowledge of the fine-grained, user-level effects of moderation interventions could better inform the content moderation process and drive the development of more effective interventions~\cite{cresci2022personalized}.

\subsection{Contributions}
Guided by these considerations, we carried out a fine-grained and multidimensional analysis of the \textit{user-level effects} experienced by core users of \subr{The\_Donald}, a Reddit community of Donald Trump supporters that was moderated by platform administrators with a sequence of interventions. Members of \subr{The\_Donald} were repeatedly denounced for toxicity, trolling, and harassment~\cite{flores2018mobilizing,massachs2020roots}. For these reasons the subreddit was quarantined in June 2019, restricted in February 2020, and finally banned in June 2020~\cite{trujillo2022make}. The quarantine removed the subreddit from the platform's search results and the feed of non-subscribed users, reducing its visibility. The restriction involved the removal of several moderators of \subr{The\_Donald} who supported content in violation of Reddit’s policies. Moreover, it allowed new submissions only from approved users, which practically halted user participation in the subreddit, causing mass user migrations to other platforms~\cite{horta2021platform,trujillo2022make}. Finally, the ban permanently shut \subr{The\_Donald} by removing it from the platform and by making it impossible for all users to access its contents. Some works already provided results about the platform and community level effects of these interventions~\cite{horta2021platform,trujillo2022make,chandrasekharan2022quarantined}. However, focusing for the first time on user-level effects allows us to seek answers to the following relevant, yet unanswered, research question:

\begin{description}
    \item[RQ:] \emph{Are user-level reactions homogeneous or heterogeneous?} In other words: \textit{Are there differences between effects at the community and user levels?}
\end{description}

% In other words: \textit{Are there significant differences in the reactions that users manifest to a moderation intervention?}
%\item \textbf{RQ1:} \textit{Are there differences between effects at the community and user levels?} In other words: \textit{Are the aggregated effects truly representative of the underlying user reactions to the moderation interventions?}  %because, with the former, we risk overlooking fringe, deviant, or outlying user reactions.

The existence of very heterogeneous user-level effects could imply that different users manifest opposite effects to the same intervention. Therefore, answering this question is important to assess the extent to which current moderation interventions are capable of producing the desired outcome on most, if not all, users. Moreover, despite being statistically infrequent and non-representative of the general behavior, fringe, deviant, and extreme reactions are those that are mostly relevant in the context of content moderation. Furthermore, exploring the relationship between effects at the community and user levels is crucial to assess the reliability of the former. In fact, depending on the distribution of user-level effects, aggregated community effects could either follow or overshadow the reactions of a minority of fringe, deviant, or extreme users. Finally, cross-checking and combining answers to the previous question opens up the possibility to explore the usefulness of user-level effects for the development of future moderation interventions, and for improving content moderation at large.

%\textbf{RQ3:} \textit{What does the knowledge of user-level effects tell us about the development of future moderation interventions and about content moderation at large?} Finally, we want to explore the implications that the knowledge of user-level effects have for content moderation. Indeed, 

\subsection{Ethics statement}
We are confident that this work will have a positive impact on the policing of online platforms by providing novel and valuable findings to inform future content moderation decisions. For example, part of our results show that certain moderation interventions result in a radicalized minority of the moderated users. In Section~\ref{sec:discussion} we discuss the ethical implications of these results, including the need to carefully balance the risk of causing harms to minorities in pursuit of benefits for the larger community. 
%This work is entirely based on public Reddit data, which was enriched with indicators of toxicity, and of the political bias and factuality of linked news articles.
%Given the sensitive information about user activities, the main ethical risk of this work is the potential deanonymization of the dataset.
This work is entirely based on public-only Reddit data, with the used dataset abiding by the FAIR principles~\cite{wilkinson2016fair}, since it is published in an interoperable and machine-readable format, and under a reusable license. Furthermore, it is indexed on the reference platform Zenodo with an associated DOI, as described in Section~\ref{sec:dataset}.

\section{Related Work}
\label{sec:related-work}
We first discuss works that assessed effects of moderation interventions at the platform and community levels, which account for the vast majority of the literature on the subject. Then, we reconsider some studies in terms of their contributions towards understanding effects at the user level.

\subsection{Effects at the platform and community levels}
%The work most similar to the present study was done by \citet{trujillo2022make}, who
In~\cite{trujillo2022make} we evaluated the community effects of the sequence of interventions on \subr{The\_Donald} ---i.e., quarantine, restriction, and ban--- finding that the first two greatly reduced the activity of the moderated users while the latter was only symbolic. However, this came at the expense of an overall trend increase in toxicity. We also concluded that the restriction had stronger effects platform-wise than the quarantine and that core users of \subr{The\_Donald} manifested more changes than the rest of users. \citet{chandrasekharan2022quarantined} evaluated quarantine effects on \subr{The\_Donald} and \subr{TheRedPill}, finding that the quarantines made it more difficult for the moderated communities to attract new members, but that the overall degree of toxicity of their existing members remained mostly unaffected. \citet{shen2022tale} studied Reddit's quarantines of \subr{The\_Donald} and \subr{ChapoTrapHouse} in terms of changes  in the activity, visibility, and political discussion of the two communities. They found that the interventions had a homogenizing effect on participation but limited effects on the visibility of community-internal issues and political language. These previous works evaluated effects at the community-level, based on the assumption that users within a community behave similarly~\cite{massachs2020roots,cresci2020emergent}. Instead, \citet{horta2021platform} evaluated effects \textit{across platforms}. They focused on users that migrated from \subr{The\_Donald} and \subr{Incels} to {\small\texttt{TheDonald.win}} and {\small\texttt{incels.co}} respectively, when the former Reddit communities got banned. Results highlighted that both bans markedly reduced user activity on the new platforms, but also that former users of \subr{The\_Donald} increased their toxicity and radicalization~\cite{horta2021platform}. Other related studies are those that evaluated the effects of moderation interventions on other Reddit communities. \citet{chandrasekharan2017you} and \citet{saleem2018aftermath} investigated the bans that targeted \subr{FatPeopleHate} and \subr{CoonTown}, uncovering that many users left Reddit after the bans, and that those who remained significantly decreased their use of hate speech~\cite{chandrasekharan2017you}. Interestingly, many \subr{CoonTown} members moved to \subr{The\_Donald} after the bans, doubling their posting activity~\cite{chandrasekharan2022quarantined}. %\citet{saleem2018aftermath} also investigated the counter-actions taken by the former \subr{FatPeopleHate} members to circumvent the ban, finding that they were short-lived and ineffective.

The effectiveness of moderation interventions was also evaluated on Instagram and Twitter. \citet{chancellor2016thyghgapp} and \citet{gerrard2018beyond} studied the effects of the 2012 ban of pro-eating disorders tags on Instagram. Results showed that, despite the intervention, the problematic content continued circulating on the platform, that users sharing such content quickly found alternative ways to identify other pro-eating disorders users, and that Instagram's recommendation system continued suggesting problematic content~\cite{gerrard2018beyond}. Moreover, pro-eating disorders communities showed increased participation, toxicity, and support for self-harm after the intervention~\cite{chancellor2016thyghgapp}. In the context of evaluating deplatforming strategies, \citet{jhaver2021evaluating} investigated the effects of Twitter's banning of the controversial influencers Alex Jones, Milo Yiannopoulos, and Owen Benjamin. They found that the intervention reduced the number of Twitter conversations about all three influencers and that their supporters exhibited decreased activity and toxicity. However, in contrast to these aggregated results, they also found that a subset of users significantly increased activity and toxicity, and measured an increased prevalence of offensive ideas and conspiracy theories associated with the banned influencers~\cite{jhaver2021evaluating}.

The above discussion reveals a broad consensus that moderation interventions tend to reduce the \textit{activity} of the moderated users~\cite{chandrasekharan2017you,saleem2018aftermath,chandrasekharan2022quarantined,trujillo2022make,horta2021platform,jhaver2021evaluating}. Regarding \textit{toxicity} however, the results are still unclear and worthy of additional investigation. While some studies measured an overall reduction in toxicity following bans~\cite{chandrasekharan2017you,jhaver2021evaluating} and other softer interventions~\cite{katsaros2022reconsidering}, others found no effects at all~\cite{chandrasekharan2022quarantined}. More worryingly, some even found increased toxicity after community bans~\cite{chancellor2016thyghgapp,horta2021platform,trujillo2022make}.
Overall, the existing results highlight that oftentimes interventions cause a mixture of desired and undesired effects.
% In addition, also in those cases when interventions produce the overall desired effects, there might be a subset of users who manifest adverse reactions~\cite{katsaros2022reconsidering}.

\subsection{Towards effects at the user level}

The previous analysis also highlights that, so far, effects of moderation interventions have been assessed almost exclusively at the level of the platform~\cite{chancellor2016thyghgapp,jhaver2021evaluating,horta2021platform} or the community~\cite{chandrasekharan2017you,saleem2018aftermath,chandrasekharan2022quarantined,trujillo2022make}. As such, we currently have very limited knowledge of user-level effects. Nonetheless, despite focusing on aggregated effects, some of the above works also provided incidental information about user-level reactions to moderation interventions. For instance, \citet{saleem2018aftermath} presented results of community-level effects as pre-post intervention scatter plots of user activity. Similarly, \citet{trujillo2021echo} presented some results as scatter plots of user activity changes. In another example, \citet{katsaros2022reconsidering} touched upon intra-user changes in toxicity, for users that received preemptive interventions on Twitter. A striking observation that arises from these studies is that user-level effects were very heterogeneous, as depicted by the large spread of points in the scatter plots in \cite{saleem2018aftermath,trujillo2021echo}. Yet, neither \citeauthor{saleem2018aftermath} nor \citeauthor{trujillo2021echo} devoted specific attention to this phenomenon, and \citeauthor{katsaros2022reconsidering} did not delve into intra-user differences. Nevertheless, these partial results imply that moderation interventions caused contrasting effects in many users. As an example, even in those cases when interventions produce the overall desired effects, there might be a subset of users who manifest adverse reactions, which has important implications for the development of future moderation interventions and for the policing of online platforms. Our present work contributes to filling this knowledge gap.

%Soft moderation interventions:~\cite{zannettou2021won}.
%Papers on \subr{The\_Donald} to cite:~\cite{flores2018mobilizing,massachs2020roots}.

\section{Dataset}
\label{sec:dataset}

% Style note: use "within and without" instead of "within and outside". The former is a fixed phrase in English that is stylistically preferable to the latter in a formal context.

For our study we utilize and enrich the dataset of core users (CUs) of \subr{The\_Donald} (TD) that we developed in~\cite{trujillo2022make}. The original dataset was obtained from Reddit's archival data on Pushshift~\cite{baumgartner2020pushshift} and is publicly available for research purposes.\footnote{\url{https://doi.org/10.5281/zenodo.6250576}} In~\cite{trujillo2022make} we defined CUs as ``those users who authored at least one post (i.e., either a submission or comment) a week, for the whole 30 weeks of the pre-quarantine period''. The dataset features 2,239 CUs and contains all of their public postings (i.e., submissions and comments) on Reddit, both within and without TD. Despite accounting for only circa 1\% of TD's users, CUs generated more than 40\% of its content before the quarantine.

\begin{figure}[t]
    \centering
    \includegraphics[width=\columnwidth]{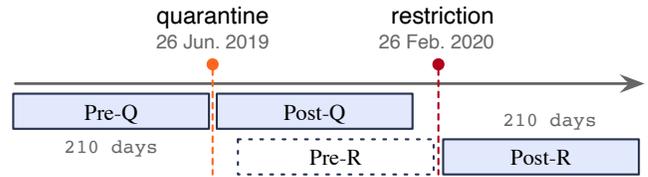}
    \caption{Timeline of the interventions on TD. Quarantine effects are assessed by comparing Pre-Q and Post-Q. Combined effects of the quarantine and restriction are assessed by comparing Pre-Q and Post-R.}
    \label{fig:periods_of_interest}
\end{figure}

Our present analyses cover the first two moderation interventions enforced on TD: the quarantine (Q) and the restriction (R). We omit analyzing the ban since TD had already been inactive for several weeks when it occurred, because of the restriction~\cite{horta2021platform}. Given that at the user level it is unwieldy to work with time series data on a daily or even weekly basis due to the high irregularity with which users create content on Reddit, we organize our data in pre-post intervention periods, as shown in Figure~\ref{fig:periods_of_interest}. We then evaluate the effects of the quarantine by comparing user behavior between the Pre-Q and Post-Q periods, and the combined effects of the quarantine and restriction by comparing user behavior between the Pre-Q and Post-R periods. For consistency with~\cite{trujillo2022make} and the definition of CUs, each period spans 210 days (30 weeks). % before or after an intervention. 
The Pre-R period is not used since it overlaps for the most part (84\%) with Post-Q. The Pre-Q, Post-Q, and Post-R periods contain respectively 3.32M, 2.51M, and 0.85M postings.

\section{Methods}
\label{sec:methods}

% Style note: "suffer a change" suggests that the change is unfavorable to the subject of the verb.

Our analyses do not aim to establish causal relationships between moderation interventions and user-level behavioral changes. Rather, we seek to describe the associations and significance between the two. Nevertheless, previous studies that used different quasi-experimental methods support the hypothesis that the interventions on TD indeed had causal effects at the community-level~\cite{horta2021platform,shen2022tale,trujillo2022make}. Naturally, these effects are the result of changes made by individual users after the interventions.

%For pre-processing and statistical computing we used the \texttt{R} ecosystem. Other than the user features described in the Dataset section, we also analyzed subreddit diversity and adopt a single indicator of change for numeric features, as described next.

%Given that many users do not have data for all of the indicators and that many reduced or ceased their activity after the interventions, when evaluating some indicators we report the number of users (n) who had data on a given period. %  or on both pre-post periods

\subsection{Characterizing user behavior}
\label{sec:behavior_indicators}

We evaluate intervention effects in terms of the changes that the quarantine and restriction caused across multiple dimensions of user behavior. In addition to the widely-studied content \textit{activity} and \textit{toxicity}, we also evaluate possible effects on the \textit{trustworthiness of the news} shared by users and on the \textit{diversity of the subreddits} in which they participate. %In the following we define the indicators used to measure these behavioral dimensions.

\noindent\textbf{Posting activity.}
We measured user-level posting activity as the number of postings (submissions and comments) published by a user in a given period.

\noindent\textbf{Comment toxicity.}
Our indicator of toxicity is based on the severe toxicity score provided by the well-known Google Perspective API~\cite{rieder2021fabrics}, which was in part trained based on Reddit comments. % from 0 to 1 of

\noindent\textbf{Trustworthiness of shared news.}
To measure the trustworthiness of the news shared by users we used the scoring of news outlets by Media Bias/Fact Check (MBFC).\footnote{\url{https://mediabiasfactcheck.com/}} More specifically, we focused on the the user-level average political bias and factual reporting level of the links shared by CUs. In detail, median scores are computed for each user, based on the links found within the postings made by that user. In MBFC, political bias is measured with an ordinal five-item scale on the US political spectrum, while factual reporting by means of six ordinal scores ranging from \emph{very low} to \emph{very high} factuality. To obtain the most representative level of users on each feature, we used the user-median on the ordinal scales. In cases in which the median was fractional, we rounded towards the user-mean level. %We thus transformed political bias into a numeric scale from -2 (left) to +2 (right), and factuality  into a scale from 1 to 6.

\begin{figure*}[t]
    \centering
    \begin{subfigure}{.47\textwidth}%
        \includegraphics[width=\textwidth]{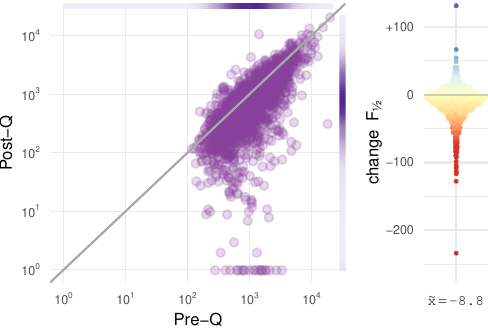}
        \caption{User activity before and after the quarantine.}\label{fig:activity_q}
    \end{subfigure}\hspace{.06\textwidth}%
    \begin{subfigure}{.47\textwidth}%
        \includegraphics[width=\textwidth]{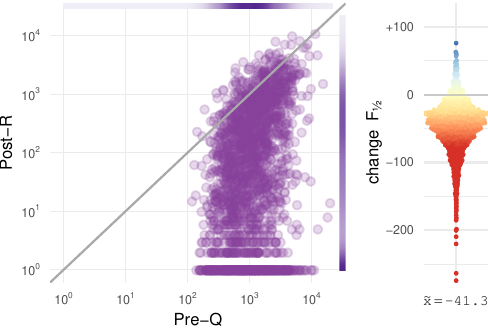}
        \caption{User activity before the quarantine and after the restriction.}\label{fig:activity_i}
    \end{subfigure}%\hspace{.1\textwidth}%
    \caption{Platform-wise activity of all users around the quarantine (a) and both interventions (b). In the scatter plots, each dot represents a user and the axes represent the number of postings in the pre-post intervention periods. Dots below the main diagonal are users who decreased their activity and those above vice versa. Users who ceased activity are squished at the bottom. Marginal distributions are shown as density strips. In the adjacent univariate beeswarm plots, each dot represents a user, positioned and colored according to the value of their activity change \Fhalf.}
    \label{fig:activity}
\end{figure*}

% Style note: Avoid "i.e.," in parenthetical phrases for alternative diversity indices; it is redundant in these cases.
\noindent\textbf{Subreddit diversity.} Subreddits represent communities with shared interests, values, and moderation practices~\cite{weld2022makes}. We are thus interested in studying if and how users changed their participation in subreddits across the platform, after the interventions. Furthermore, measuring diversity in participation is important, since lack of diversity is linked to the emergence of echo chambers~\cite{matakos2020maximizing}. We measure subreddit participation diversity in a given period via the Hill diversity index~\cite{hill1973diversity}, which extends and unifies various metrics traditionally used for diversity, including richness (the mere count of types), Shannon index (a measurement of entropy), and Gini-Simpson index (a probability). We adapt the Hill diversity to measure user participation in subreddits as:

\begin{equation*}
    ^qD = \left( \sum_{i=1}^S p_i^q \right)^{1/1-q}    
\end{equation*}

\noindent where $D$ is diversity; $q$ is the order of the diversity (increasing $q$ generally results in more weight given to abundant subreddits); $S$ is the number of distinct subreddits (richness); and $p_i$ is the relative abundance of subreddit $i$. When $q=1$, the index $^1D$ is called Hill-Shannon diversity because it is linked to the Shannon index $H'$, as $H'= \ln({^1}D)$. Herein we use $^1D$ given its balance between rare and abundant subreddits, with values ranging from $1$ when a given CU participates exclusively in a single subreddit, to richness $S$ when there is an equal proportion among subreddits in which a user participates. We remark that our data allows measuring subreddit diversity only in terms of active user participation (i.e., posting submissions and comments). As such, some users exhibiting low diversity might still passively browse many subreddits.

\subsection{Quantifying effects}
When presenting results, we use the median ($\tilde{x}$) to indicate the central tendency of a distribution; the median absolute deviation (\textsc{mad}) to indicate the spread; Kendall's $\tau$ coefficient for association significance between two dimensions; one-sided Wilcoxon signed-rank test ($V$) for paired data (of the same user) before and after intervention; and Wilcoxon-Mann-Whitney test for group independence ($Z$) between different groups of users. For the figures, we aimed to display all individual users while making evident the outliers, as these individuals can greatly influence the value of an indicator when measured at the community level. Naturally, this also extends to intervention effects, which we define as changes in a dimension of user behavior between pre-post intervention periods.

\noindent\textbf{Indicator of change.} Changes in numeric variables are usually measured either in absolute or relative terms. However, both have important limitations since absolute change by itself is seldom useful without a value of reference, whereas relative change ignores the magnitude of change and it is not antisymetric. Logarithmic differences are an antisymetric alternative, but they also ignore the magnitude of the change. For these reasons, usually both absolute change and relative change (or log differences) are used as indicators of change. Here, we utilize a single change indicator recently proposed by \citet{brauen2020absolute}, which takes into account both magnitude and relative differences: the function $F_{\lambda}(a,b)$, with $\lambda \in [0,1]$, a unitless antisymetric indicator of the change experienced by a variable $x \in \mathbb{R}$ when passing from value $a$ to $b$. It is defined as:
%Recently, a family of change indicators that take into account both magnitude and relative differences has been proposed \cite{brauen2020absolute}, which are related to both relative change and log differences.  

\begin{equation*}
    F_{\lambda}(a,b) = \begin{cases}
        \cfrac{b^{1-\lambda}-a^{1-\lambda}}{1-\lambda} & \text{if} \, \ \lambda \neq 1 \\
        \ln(b/a) & \text{if} \, \ \lambda = 1 
    \end{cases}
\end{equation*}

$F_{\lambda}(a,b)$ interpolates between absolute change ($\lambda = 0$) and logarithmic differences ($\lambda = 1$). For our analyses we use $\lambda = \text{½}$ in order to interpolate midway between the two, with the indicator of change $F_{\text{½}}$ used herein being:

\begin{equation*}
    F_{\text{½}}(a,b) =  \cfrac{\sqrt{b} - \sqrt{a}}{\text{½}}
\end{equation*}

\section{Results}
\label{sec:results}
In the following subsections we present the results regarding the change of posting activity, subreddit diversity, and comment toxicity, as well as a description of the sharing of news links in terms of factual reporting and political bias. In addition, based on the large decrease in posting activity, we also delve into an analysis of user account inactivation, which is the ceasing of all platform-wise activity following the quarantine and restriction. Then, we discuss these results and their implications in Section~\ref{sec:discussion}.

Since many users reduced or ceased their activity after the interventions, or never shared a link to a news outlet present in MBFC, we could not compute all the indicators for every user and every period. Hence, when illustrating results for some indicators, we report the number of active users ($n$) involved in the analysis.

\subsection{Posting activity}
In aggregate, the median user activity before the quarantine (Pre-Q) was 1,051 postings (\madval{833}). Median user activity decreased to 711 (\madval{682}) in Post-Q and dropped to 51 (\madval{76}) in Post-R, demonstrating the effectiveness of the two interventions at reducing user activity, in agreement with previous community-level results~\cite{trujillo2022make}.
When analyzing effects at the user level, we see that the majority of users (72\%) decreased their activity after the quarantine, with a median change \Fhalf{} = -8.8 (\madval{15.8}), corresponding to a median for absolute change of -217 (\madval{438}) and for relative change of -0.26\% (\madval{0.41}). Some users ($n=18$) stopped posting altogether, as shown in Figure~\ref{fig:activity_q}. Some even manifested extreme activity changes. The user with the highest increase (\Fhalfval{131}) went from 1.4k postings in Pre-Q to 10.6k in Post-Q, whereas the one with the highest decrease (\Fhalfval{-239}) went from 18k to 311.

When considering both interventions (Figure~\ref{fig:activity_i}), 89\% of the users decreased their activity, as can be seen by the remarkable drop in \Fhalf{} (\medmadval{-41.3}{26}). Moreover, a notable number of users ($n=407$) ceased activity in Post-R. The beeswarm plot of Figure~\ref{fig:activity_i} shows the distribution of user-level activity changes. In comparison with that of Figure~\ref{fig:activity_q}, we see that the sequence of both interventions had stronger effects (\medval{-41.3} \textit{vs} \medval{-8.8}) than the quarantine alone. Interestingly, we also note that the distribution of \Fhalf{} is much more spread out (\madval{26} \textit{vs} \madval{15.8}), with many users exhibiting important changes in activity (both decreases and increases). In addition, although most users were consistent in their direction of change, some manifested contrasting changes. In detail, 42 users decreased activity after the quarantine, but increased it after the restriction, whereas 525 did the opposite. Finally, we measured no correlation between account age and activity, or change in activity, for each intervention, with $\tau$ being close to 0 and \mbox{$p>$ .28} in all cases.

\subsection{Subreddit diversity}
Overall, we measured a low median subreddit diversity in Pre-Q (\medmadval{2.8}{2.7}), meaning that the majority of users showed a strong preference for a very limited number of subreddits. As an example, 123 users (5\% of the total) participated exclusively in TD during this period. These findings might support the existence of political echo chambers. As such, they appear to contradict previous studies that reported no evidence of echo chambers among Reddit supporters of Donald Trump~\cite{de2021no}. We note however that conclusive results on this regard would mandate a detailed analysis of the aforementioned accounts, which could simply be bogus or throwaway accounts specifically created by some users to remain anonymous within TD.

\begin{figure}[t]
    \centering
    \includegraphics[width=\columnwidth]{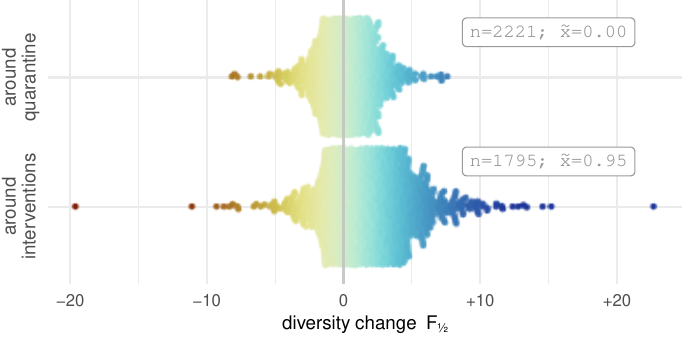}
    \caption{Change in subreddit diversity for active users. After both interventions, remaining users participated more in other subreddits (in many cases considerably much more). }\label{fig:diversity_change}
\end{figure}

We measured a weak, yet significant, positive correlation between subreddit diversity and posting activity ($\tau$\,=\,.18; p\,$\ll$\,.01), as well as between diversity and account age ($\tau$\,=\,.12; p\,$\ll$\,.01). Interestingly, we note that the diversity for users who ceased activity after the restriction was \textit{significantly lower} (Z=-11.10; p\,$\ll$\,.01) with respect to that of the users who kept on posting on Reddit. In other words, the less diverse a user is in their subreddit participation habits, the more likely they are to stop activity on Reddit after the restriction. With reference to the 123 users who participated only in TD, 56 of them (46\%) ceased activity in Post-R.

User-level changes (\Fhalf{}) in subreddit diversity can be computed only for those users who stayed active after one or both moderation interventions. For active users after the quarantine, we measured a balanced change in diversity (\medmadval{0}{0.54}), as shown in the top row of Figure~\ref{fig:diversity_change}. When considering both interventions we found a moderately positive (\medmadval{.94}{1.86}) change in diversity, meaning that after the restriction active users participated in an increased number of subreddits. As visible from the bottom row of Figure~\ref{fig:diversity_change}, the majority of outliers also had positive changes. Indeed, the overall median diversity increased from 2.8 in Pre-Q to 6.7 (\madval{7.4}) in Post-R. These results are consistent with work at the community level \cite[\S 5.3.5]{trujillo2022make}. 

\subsection{Comment toxicity}
\begin{figure}[t]
    \centering
    \includegraphics[width=\columnwidth]{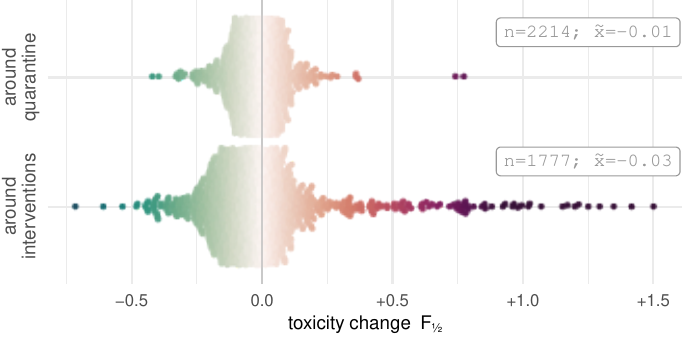}
    \caption{Change in comment toxicity for active users. Despite still being balanced in direction, after both interventions the change was asymmetrical, with many users remarkably increasing their toxicity.}\label{fig:toxicity_change}
\end{figure}

In Pre-Q, the median user toxicity score was .060 (\madval{.02}), which decreased slightly but steadily both in Post-Q (\medmadval{.058}{.02}) and Post-R (\medmadval{.052}{.02}). As shown in the top row of Figure~\ref{fig:toxicity_change}, user-level changes in toxicity after the quarantine were mostly symmetrical and concentrated in the region of \Fhalf{} = 0, meaning that the vast majority of users only manifested minor changes and that users who increased their toxicity were counterbalanced by a similar number of users who decreased it. Nonetheless, the figure also shows a couple of outlier users (dark-colored) who increased their toxicity substantially (\Fhalf{} $\simeq$ 0.75).

\begin{figure*}[t]
    \centering
    \includegraphics[width=\textwidth]{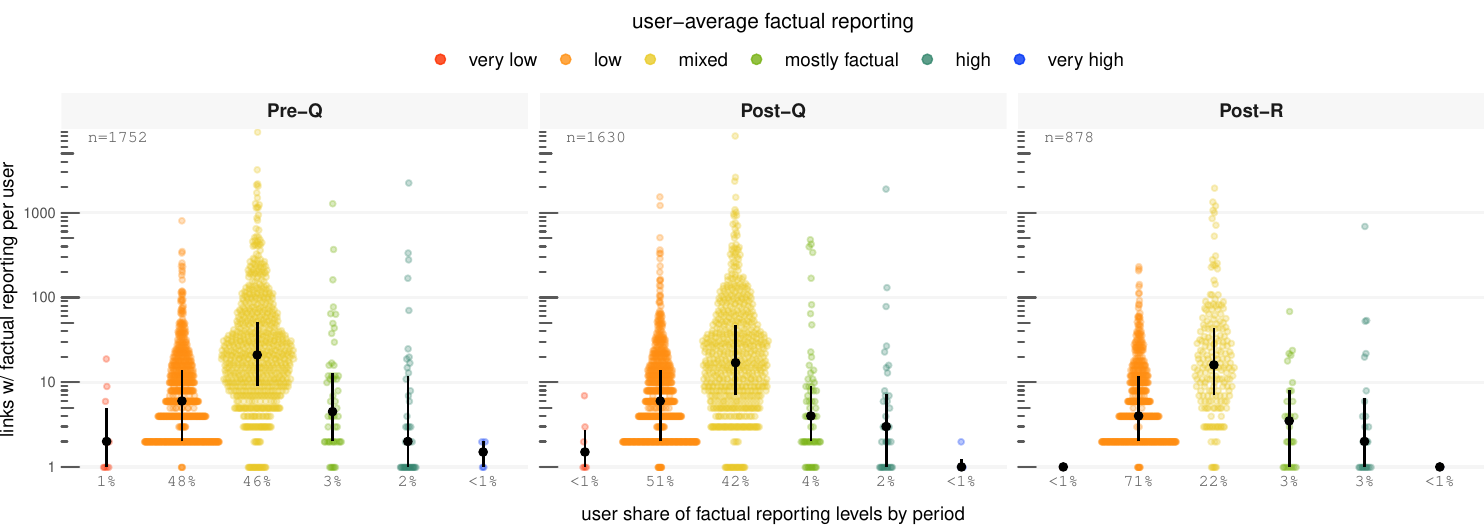}
    \caption{User-average factual reporting level of shared news outlets for active users. Each dot represents a user, colored based on their most representative level and positioned according to the number of news links shared. Percent values represent the proportion of users with an average factuality level for a given period. Medians and interquartile ranges are shown in black.}\label{fig:factual_reporting}
\end{figure*}

Conversely, considering both interventions surfaces important differences between the effects manifested by the majority of the users and those manifested by the outliers. In fact, the majority of users slightly decreased their toxicity after the restriction, as demonstrated by a median \Fhalf{} = -0.03. At the same time, the bottom row of Figure~\ref{fig:toxicity_change} shows that a significant number of users diverged from the bulk of the distribution. The majority of such divergent users manifested strong increases in toxicity, while a minority manifested moderate decreases. In other words, this result highlights that, when evaluated at the community level, the restriction caused a slight toxicity decrease. However, the user-level analysis reveals that a significant number of users \textit{strongly increased their toxicity}, in spite of the opposite community effect. Incidentally, at the community level there was a surge in toxicity around the beginning of the George Floyd protests \cite[\S 5.3.2]{trujillo2022make}, thus it is likely that the outliers who increased their toxicity did so due to these events. This is an example of an exogenous event that can render more challenging the study of causality between intervention effects and changes in user behavior. There was no significant correlation between the number of comments and the toxicity of users in any of the three periods ($\tau$\,$\approx$\,.01; p\,$>$\,.14). In Pre-Q, there was a very weak significant negative correlation between subreddit diversity and toxicity ($\tau$\,=\,-0.09; p\,$\ll$\,.01), meaning that users with less diverse subreddit participation habits had a slight tendency to be more toxic. Table~\ref{tab:change_summary} contains a summary of the change indicators of user-level posting activity, subreddit diversity, and comment toxicity.

\begin{table}
    \centering
    \small
    \begin{tabular}{lcrrcrr}
        \toprule
        && \multicolumn{5}{c}{\textbf{periods}} \\
        \cmidrule{3-7}
        && \multicolumn{2}{c}{Pre-Q / Post-Q} && \multicolumn{2}{c}{Pre-Q / Post-R} \\
        \cmidrule{3-7}
        && $\tilde{x}$ & \textsc{mad} && $\tilde{x}$ & \textsc{mad} \\
        \midrule
        posting activity && -8.8 & 15.8 && -41.3 & 25.9 \\
        subreddit diversity && +.005 & .535 && +.946 & 1.856 \\
        comment toxicity && -.006 & .037 && -.026 & .079 \\
        \bottomrule
    \end{tabular}
    \caption{User-level median and spread values of behavior change indicators \Fhalf{}.}
    \label{tab:change_summary}
\end{table}

\subsection{Trustworthiness of shared news}
For the three periods of interest (Pre-Q, Post-Q, and Post-R) there was a total of 372k submissions. Circa 220k submissions had an external link (i.e., pointing outside of Reddit), with 23k of links pointing to a news outlet contained in the MBFC repository. The vast majority of these links pointed to news outlets labeled as questionable sources (64\%) or as politically biased (32\%). The rest of the links (4\%) pointed to news outlets classified as either satire, conspiracy/pseudoscience, or pro-science.

To investigate intervention effects on the factuality of the shared news, we associated each user to a news factual reporting score from MBFC. For each user, the factuality score is computed as the median of the factuality scores of the news outlets that the user linked in their submissions. Figure~\ref{fig:factual_reporting} shows, for each period, the relationship between user factuality scores and the number of shared links. This analysis allows evaluating whether users sharing more or less factual news are more or less prolific than others. In addition, it also allows assessing whether the moderation interventions on TD altered this relationship in some way. Throughout all three periods the most common user factuality score was \textit{low}, as shown in Figure~\ref{fig:factual_reporting} (orange-colored distributions). Users sharing \textit{low} factuality news accounted for 48\% of all users in Pre-Q, 51\% in Post-Q, and 71\% in Post-R, demonstrating a steady increase. The second most common factuality score was \textit{mixed}, which went from 46\% in Pre-Q, to 42\% in Post-Q, and 22\% in Post-R. The combination of the remaining scores accounted for \textless 7\% of users in all three periods. This analysis highlights that the sequence of interventions on TD resulted in a remarkable fraction of users sharing relatively less factual news, particularly because many users who had an average  \textit{mixed} factuality passed to an average \textit{low} factuality, with a respective difference of –24 and +23 percent points. When considering the number of shared links per user, we observe that the most prolific users are those with a \textit{mixed} factuality score. This is reflected in Figure~\ref{fig:factual_reporting} by the spread and the outliers of the yellow-colored distributions. This phenomenon is mostly visible in Post-R, where the top-5 link sharers (representing 0.6\% of the 878 active users) all had \textit{mixed} factuality and published more than a thousand posts each, producing 28\% of all the links shared in that period.

\begin{figure*}[t]
    \centering
    \includegraphics[width=\textwidth]{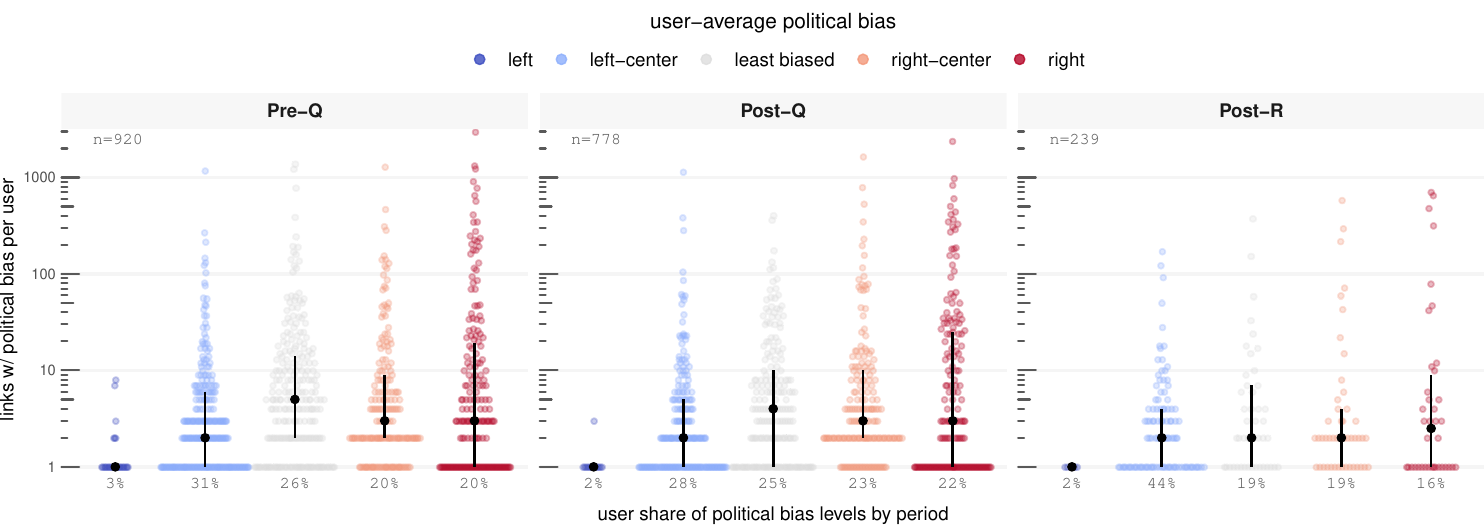}
    \caption{User-average political bias level of shared news outlets for active users. Each dot represents a user, colored based on their most representative level and positioned according to the number of news links shared. Percent values represent the proportion of users with an average bias level for a given period. Medians and interquartile ranges are shown in black.}\label{fig:political_bias}
\end{figure*}

\begin{table}[t!]
    \centering
    \small
    \begin{tabular}{lcrrr}
        \toprule
        && \multicolumn{3}{c}{\textbf{link share (\%)}} \\
        \cmidrule{3-5}
        \textbf{user-average bias} && Pre-Q & Post-Q & Post-R \\
        \midrule
        left         &&  1.6 &  0.2 &  0.9 \\
        left-center  && 16.2 & 14.5 & 16.2 \\
        least biased && 32.3 & 19.6 & 14.1 \\
        right-center && 18.0 & 28.8 & 34.8 \\
        right        && 31.9 & 36.9 & 34.0 \\
        \bottomrule
    \end{tabular}
    \caption{Link share of politically biased news. For each successive period the share of links by users who publish mainly right-center and right leaning outlets increases.}
    \label{tab:pol_bias_link_share}
\end{table}

We repeated this analysis also for the political bias of the shared news. Perhaps surprisingly, Figure~\ref{fig:political_bias} shows that the majority of users had a median \textit{left-center} (light-blue) political bias. They accounted for 31\% of all users in Pre-Q, 28\% in Post-Q, and 44\% in Post-R. Regarding extremely biased users, we note a strong prevalence of \textit{right} biased users in all three periods, accounting for 20\% of all users in Pre-Q, 22\% in Post-Q, and 16\% in Post-R. \textit{Left} biased users always accounted for $\le$ 3\% of all users. The analysis of the most prolific users reveals an interesting trend. In all three periods, the users that shared the largest number of links always laid at the right side of the political spectrum, as shown by the fat tails, and by the presence of many outliers, in the distributions of the user with \textit{right-center} and \textit{right} political bias.
As already observed for factual reporting, this phenomenon is particularly prevalent in Post-R. As visible from the rightmost panel of Figure~\ref{fig:political_bias}, the more biased towards the right a user is, the more prolific they are. As an example, the top-2 users who shared the most links in Post-R, respectively 696 and 643 links, are \textit{right} biased and accounted for 25\% of all links shared in that period. This phenomenon is also visible in Table~\ref{tab:pol_bias_link_share}, which reports the percentage of links shared for each class of political bias, in each period. Notably, the sum of the \textit{right-center} and \textit{right} biased links accounted for 49.9\% of all links in Pre-Q, 65.7\% in Post-Q, up to 68.8\% in Post-R, demonstrating that the moderation interventions on TD caused a progressive polarization of the affected users. The percentage of \textit{left-center} and \textit{left} biased links remained roughly the same throughout the three interventions, while \textit{least biased} links decreased steadily.

\subsection{User account inactivation}
\begin{figure}[t]
    \centering
    \includegraphics[width=\columnwidth]{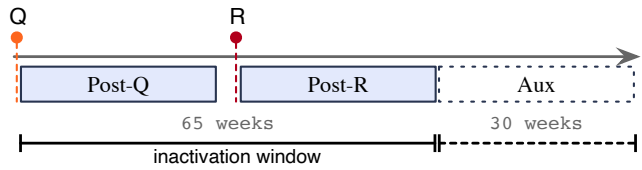}
    \caption{We marked as \emph{inactivations} those users who last posted content within the 65 weeks after the quarantine (Q), taking also into account activity within an auxiliary period of 30 weeks beyond the period after restriction (R).}\label{fig:inactivation_time_window}
\end{figure}

Due to the important reduction in active users, we delved into the effects of the interventions on user account inactivations. To this end, we collected additional data corresponding to the CUs activity during an auxiliary period covering 30 weeks following the end of Post-R (i.e., 65 weeks after quarantine). We then derived the last date in which a user published a posting in Reddit during this extended time frame of 95 weeks. Finally, we defined as \textit{inactive users} those whose last posting date was within the 65 weeks following the quarantine, up to the end of the Post-R period, as depicted in Figure~\ref{fig:inactivation_time_window}.

At the platform level, we identified three kinds of user account inactivations: abandoned, deleted, and suspended. In an \emph{abandoned} account a user simply stopped posting content to the platform. In a \emph{deleted} account the user deliberately inactivated their account via the platform. In that case it can’t be reactivated, their username becomes unavailable, and they lose access to their account and posting history, and all postings are disassociated from the user but remain on the platform. If the user would like to delete the contents of the postings, they would need to do so prior to account deletion. In a \emph{suspended} account, a Reddit administrator has forcefully shut the account, following violations of the platform's policies. To derive the account status of all CUs, we used the official Reddit API. %It should be noted that deleted accounts only disassociate postings on Reddit itself; for the archival data present in Pushshift the user would need to make an additional deletion request to the archive.
% The suspension may be permanent or temporary; given that suspended core users ceased all activity we presumed that it is permanent. 

\begin{figure}
    \centering
    \includegraphics[width=\columnwidth]{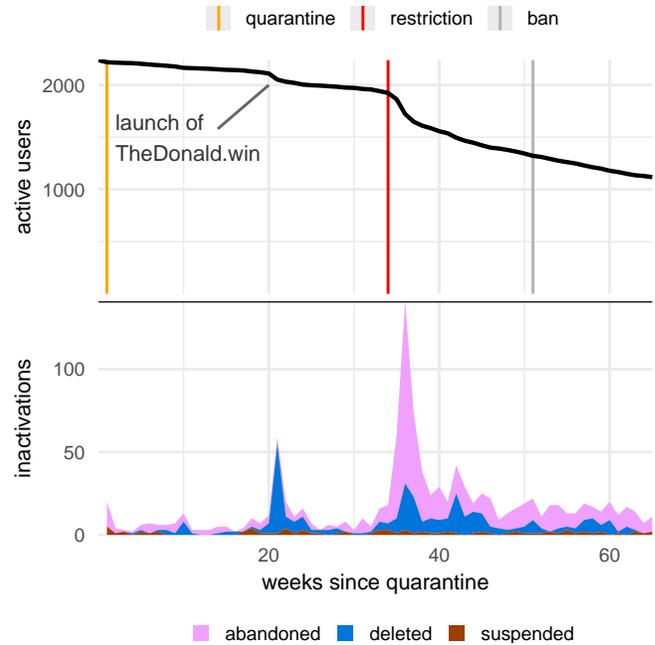}
    \caption{Time series of remaining active core users (top line chart) and corresponding user account inactivations (bottom area chart) after the quarantine and restriction. For completeness, we also include the ban, which did not have a visible effect on inactivations, unlike the launch of the forum TheDonald.win by former members of the subreddit. }\label{fig:inactivations}
\end{figure}

During the 65 weeks after quarantine, there were 1,121 account inactivations, slightly more than half of the initial core users, with 691 (62\%) being abandoned, 348 (31\%) deleted, and 82 (7\%) suspended. As shown in Figure~\ref{fig:inactivations}, most of the inactivations (62\%) occurred within the 30 weeks of the Post-R period, mostly in the few weeks after the restriction. Regarding the Post-Q period, interestingly most inactivations occurred after the launch of the forum {\small\texttt{thedonald.win}} by former members of TD, most likely due to the migration of many of the subreddit users to the newly created platform~\cite{horta2021platform}. In addition, in contrast to the trend of most of the time frame, upon launch of the forum, most inactivations were in the form of deleted accounts instead of abandoned ones, with 43\% and 39\% respectively in the following month.

During the Pre-Q period, inactivated users published less postings, both by total of postings (1.59M \textit{vs} 1.73M) and by user-level median (980 \textit{vs} 1,109), with the latter being a significant difference ($Z$\,=\,-3.84; $p<$\,.01). Concerning subreddit diversity, inactivated users were also less diverse (\medmadval{1.73}{1.08}) compared to remaining users (\medmadval{4.73}{4.86}). The difference is statistically significant ($Z$\,=\,-17.3; $p\ll$\,.01). For comment toxicity, on the other hand, there was significant higher toxicity ($Z$\,=\,6; $p\ll$\,.01) among inactivated users (\medmadval{.063}{.02}) compared to the remaining active users (\medmadval{.057}{.02}). With regards to the trustworthiness of shared news, we used the Cochran–Armitage test ($Z_C$) to check for significant differences in user-average factual reporting and political bias by inactivation status. In general, remaining active users shared content leaning to the left of the US political spectrum compared to inactivated users ($Z_C$\,=\,-3.1, $p<$\,.01). At the same time, remaining users shared news from sources with lower factuality compared to inactivated users ($Z_C$\,=\,-2.8, $p<$\,.01).

% Statistical note: we did not use polyserial correlation to compare the number of links with their quality levels because the underlying joint distribution is not bivariate normal.

\section{Discussion}
\label{sec:discussion}
%\hl{In this section we talk about fat tailed distributions of effects: but are they really fat tailed? Maybe they are still normally distributed, but with a larger deviation?}

Our analyses provide novel and nuanced insights into the effects that the quarantine and the restriction had on the core users of \subr{The\_Donald}. Overall, the two interventions had comparable effects, although with different magnitudes. In particular, at the user level both the quarantine and the restriction: (\textit{i}) strongly reduced user activity, (\textit{ii}) slightly increased the diversity of the subreddits in which users participated, (\textit{iii}) very slightly reduced user toxicity, and (\textit{iv}) led users to share less factual and more politically biased news, especially towards the right side of the political spectrum. For each of these effects, we found that the restriction had a stronger impact than the quarantine, as demonstrated by the larger median values reported in Table~\ref{tab:change_summary} for effects of both interventions (Pre-Q/Post-R) with respect to those of the quarantine (Pre-Q/Post-Q). These user-level results mostly confirm previous ones obtained at the community level for the same moderation interventions~\cite{horta2021platform,chandrasekharan2022quarantined,trujillo2022make,shen2022tale}. In addition, our analyses also allow to provide answers to our RQ.
%\hl{Can we include a table that summarizes the user level median and MAD for each intervention and for each indicator of user behavior? The same table could also have a column to recap community level effects for the same indicators, so as to allow a comparison of the user and community level effects.} 

\noindent\textbf{Community versus user-level effects.} %In addition, our analyses also allow to investigate the fine-grained user level dynamics that led to the emergence of the community-level effects reported in previous works.
An interesting finding of our work is that, for each intervention and for each dimension of user behavior, there were outliers who manifested exaggerated effects and that significantly deviated from the average community reactions. In Table~\ref{tab:change_summary} this is reflected by \textsc{mad} values that are larger than the corresponding median, in all but one case (i.e., restriction effect on user activity). The fact that a minority of outliers manifested effects that were several times stronger than those of the other users, implies that aggregated community level effects can be strongly influenced by the behavior of a minority of users. This mandates care when evaluating intervention effects exclusively at the platform or community level, as done in the majority of existing works~\cite{chancellor2016thyghgapp,jhaver2021evaluating,horta2021platform,chandrasekharan2017you,saleem2018aftermath,chandrasekharan2022quarantined,trujillo2022make}. In fact, depending on the underlying distribution of user-level effects, aggregated intervention effects might be weakly representative of the general user-level behaviors, being strongly dependent on the behavior of a handful of outliers. %\hl{In our data, this is reflected by \textsc{mad} values that are always larger than the median.}
Or conversely, community effects could overshadow and conceal the behavior of some user minorities. The former scenario is well documented in the literature about online harms, where it is typical for a minority of fringe users to be responsible for the majority of the harms~\cite{zannettou2020measuring,robertson2022uncommon}. Instead, the latter possibility relates to \citeauthor{welles2014minorities}' case for \textit{making Big Data small}, in that platform and community effects could ``silence [minorities and outliers] through statistical aggregation''~\cite{welles2014minorities}. In any case, our results suggest that increased attention should be devoted to the user-level effects of moderation interventions, and demand additional efforts for their measurement.

\noindent\textbf{Heterogeneous user-level effects.} Our results also highlighted that the presence of outliers and fat tailed or highly skewed distributions of user-level effects were more prominent after the restriction on \subr{The\_Donald} and less so after the quarantine. In our quantitative results, this is demonstrated by larger \textsc{mad} values for the effects of the restriction than for those of the quarantine, as reported in Table~\ref{tab:change_summary}. Combined with previous findings, this important result tells us that the restriction had stronger effects overall, but also that it was associated with \textit{more extreme and diversified user reactions}, perhaps also in response to exceptional events in this tumultuous political period, such as the George Floyd protests~\cite{trujillo2022make}. %A paramount example are the effects that the restriction had on user toxicity. The intervention slightly reduced the toxicity of the community as a whole, but at the same time it caused a subset of users to become extremely toxic, as demonstrated by the long tail of positive user toxicity changes shown in the bottom row of Figure~\ref{fig:toxicity_change}.
If confirmed in other contexts, this finding can have important practical and ethical implications for the policing of online platforms. Indeed, the choice of a moderation intervention can affect the balance between the effectiveness of the intervention at large, and the extreme ---possibly undesired--- deviations it might cause to the behavior of some users. A prime example in our work are the different rates at which users ceased to use their account (be it for abandonment, deletion, or suspension) after interventions or external events linked to the interventions, such as the creation of a more polarized or less moderated alternative online space. For the future, moderators and platform administrators should be aware of this issue and should account for both community and user level reactions when deciding on the enforcement of a moderation intervention. From an ethical standpoint, our results also call for renewed attention on the delicate balance between \textit{common} versus \textit{minority} good~\cite{velasquez1992common}. %\hl{Can we find a better reference?}
Scholars and practitioners are now faced with the question as to whether it is right to risk causing serious harm to a minority of deviant users, in order to obtain a mild benefit for the larger community. %Importantly, the need to balance out these contrasting effects also opens new directions of research, for example with the goal of designing new moderation interventions that are capable of reaching community goals without sacrificing those of the individual users.

\noindent\textbf{Future work in content moderation.} The results presented herein also have important implications for the design and deployment of future moderation interventions. Specifically, we showed that each intervention, independently of the type and magnitude of its effects, was associated to \textit{diverse user reactions}. In other words, different users reacted differently to the same interventions. Related literature also showed that applying the same intervention to the same users multiple times, was also linked to diverse (e.g., reduced) effects~\cite{katsaros2022reconsidering}. The existence of these heterogeneous reactions, which we measured empirically, is consistent with relevant theories from the social and cognitive sciences. These posit that user reactions to moderation interventions and other persuasive efforts are based on each user's individual characteristics and on the context of the moderation~\cite{williams2017individual,molina2022does}. Both these theories and our present results suggest that it is unlikely for a single moderation intervention to produce the desired effects (e.g., toxicity reduction) for \textit{all} moderated users. On the contrary, developing \textit{targeted} interventions that account for individual and contextual characteristics could lead to more effective and user-centered content moderation processes. For example, with respect to our previous discussion on minorities and outliers, future research could aim at designing moderation interventions that are capable of reaching community goals without sacrificing those of the individual users. To this end, our results support the recent experimentation with diversified moderation interventions~\cite{bilewicz2021artificial} and the proposal of \textit{personalized content moderation}~\cite{cresci2022personalized}.

\section{Conclusions}
\label{sec:conclusions}
We evaluated the effects that the quarantine and the restriction had on the core users of \subr{The\_Donald}. Differently from previous work, we assessed effects at the \textit{user level}, finding that the interventions produced some of the intended outcomes, including the reduction of user activity and toxicity, and the increase in the diversity of the subreddits in which users participated. However, both interventions also produced some unintended effects, as they led remaining users to proportionally share less factual and more politically biased news. Our results also highlighted that the interventions that appeared as overall more effective (i.e., that produced stronger effects) were also associated with more varied user reactions. We conclude that platform and community level effects are not always representative of the underlying behavior of individuals or smaller user groups. We discussed the practical and ethical implications of our findings, which hopefully motivate future research and experimentation on diversified and personalized content moderation. In addition, the existence of very diverse user reactions bears the question as to why such differences exist in the first place, and what are the characteristics that differentiate users who react differently. A limitation of our work is the focus on core users who were actively involved in a subreddit for a prolonged time (and who produced the most content), but which excludes users with less prominent participation levels (and the respective intervention effects on these). In addition, our data features matched samples of users in multiple intervention periods, which lends itself to more sophisticated analyses with mixed models. Focusing on these aspects represents a promising direction for future work and experimentation.

\bibliographystyle{ACM-Reference-Format}
\bibliography{references}


\begin{thebibliography}{37}

%%% the Web address.

\ifx \showCODEN    \undefined \def \showCODEN     #1{\unskip}     \fi
\ifx \showDOI      \undefined \def \showDOI       #1{#1}\fi
\ifx \showISBNx    \undefined \def \showISBNx     #1{\unskip}     \fi
\ifx \showISBNxiii \undefined \def \showISBNxiii  #1{\unskip}     \fi
\ifx \showISSN     \undefined \def \showISSN      #1{\unskip}     \fi
\ifx \showLCCN     \undefined \def \showLCCN      #1{\unskip}     \fi
\ifx \shownote     \undefined \def \shownote      #1{#1}          \fi
\ifx \showarticletitle \undefined \def \showarticletitle #1{#1}   \fi
\ifx \showURL      \undefined \def \showURL       {\relax}        \fi
% invisible to TeX
\providecommand\bibfield[2]{#2}
\providecommand\bibinfo[2]{#2}
\providecommand\natexlab[1]{#1}
\providecommand\showeprint[2][]{arXiv:#2}

\bibitem[Andre and Velasquez(1992)]%
        {velasquez1992common}
\bibfield{author}{\bibinfo{person}{Claire Andre} {and} \bibinfo{person}{Manuel
  Velasquez}.} \bibinfo{year}{1992}\natexlab{}.
\newblock \showarticletitle{The common good}.
\newblock \bibinfo{journal}{\emph{Issues in Ethics}} \bibinfo{volume}{5},
  \bibinfo{number}{2} (\bibinfo{year}{1992}).
\newblock


\bibitem[Baumgartner et~al\mbox{.}(2020)]%
        {baumgartner2020pushshift}
\bibfield{author}{\bibinfo{person}{Jason Baumgartner}, \bibinfo{person}{Savvas
  Zannettou}, \bibinfo{person}{Brian Keegan}, \bibinfo{person}{Megan Squire},
  {and} \bibinfo{person}{Jeremy Blackburn}.} \bibinfo{year}{2020}\natexlab{}.
\newblock \showarticletitle{The {Pushshift Reddit} dataset}. In
  \bibinfo{booktitle}{\emph{AAAI ICWSM}}.
\newblock


\bibitem[Bilewicz et~al\mbox{.}(2021)]%
        {bilewicz2021artificial}
\bibfield{author}{\bibinfo{person}{Michal Bilewicz}, \bibinfo{person}{Patrycja
  Tempska}, \bibinfo{person}{Gniewosz Leliwa}, \bibinfo{person}{Maria
  Dowgiallo}, \bibinfo{person}{Michalina Tanska}, \bibinfo{person}{Rafal
  Urbaniak}, {and} \bibinfo{person}{Michal Wroczynski}.}
  \bibinfo{year}{2021}\natexlab{}.
\newblock \showarticletitle{Artificial intelligence against hate: Intervention
  reducing verbal aggression in the social network environment}.
\newblock \bibinfo{journal}{\emph{Aggressive Behavior}} \bibinfo{volume}{47},
  \bibinfo{number}{3} (\bibinfo{year}{2021}).
\newblock


\bibitem[Blaya(2019)]%
        {blaya2019cyberhate}
\bibfield{author}{\bibinfo{person}{Catherine Blaya}.}
  \bibinfo{year}{2019}\natexlab{}.
\newblock \showarticletitle{Cyberhate: A review and content analysis of
  intervention strategies}.
\newblock \bibinfo{journal}{\emph{Aggression and Violent Behavior}}
  \bibinfo{volume}{45} (\bibinfo{year}{2019}).
\newblock


\bibitem[Brauen et~al\mbox{.}(2020)]%
        {brauen2020absolute}
\bibfield{author}{\bibinfo{person}{Silvan Brauen}, \bibinfo{person}{Philipp
  Erpf}, {and} \bibinfo{person}{Micha Wasem}.} \bibinfo{year}{2020}\natexlab{}.
\newblock \bibinfo{title}{On absolute and relative change}.
\newblock
\newblock
\showeprint[arxiv]{2011.14807}


\bibitem[Chancellor et~al\mbox{.}(2016)]%
        {chancellor2016thyghgapp}
\bibfield{author}{\bibinfo{person}{Stevie Chancellor},
  \bibinfo{person}{Jessica~Annette Pater}, \bibinfo{person}{Trustin Clear},
  \bibinfo{person}{Eric Gilbert}, {and} \bibinfo{person}{Munmun De~Choudhury}.}
  \bibinfo{year}{2016}\natexlab{}.
\newblock \showarticletitle{\#thyghgapp: {Instagram} content moderation and
  lexical variation in pro-eating disorder communities}. In
  \bibinfo{booktitle}{\emph{ACM CSCW}}.
\newblock


\bibitem[Chandrasekharan et~al\mbox{.}(2022)]%
        {chandrasekharan2022quarantined}
\bibfield{author}{\bibinfo{person}{Eshwar Chandrasekharan},
  \bibinfo{person}{Shagun Jhaver}, \bibinfo{person}{Amy Bruckman}, {and}
  \bibinfo{person}{Eric Gilbert}.} \bibinfo{year}{2022}\natexlab{}.
\newblock \showarticletitle{Quarantined! Examining the effects of a
  community-wide moderation intervention on {Reddit}}.
\newblock \bibinfo{journal}{\emph{ACM TOCHI}} \bibinfo{volume}{29},
  \bibinfo{number}{4} (\bibinfo{year}{2022}).
\newblock


\bibitem[Chandrasekharan et~al\mbox{.}(2017)]%
        {chandrasekharan2017you}
\bibfield{author}{\bibinfo{person}{Eshwar Chandrasekharan},
  \bibinfo{person}{Umashanthi Pavalanathan}, \bibinfo{person}{Anirudh
  Srinivasan}, \bibinfo{person}{Adam Glynn}, \bibinfo{person}{Jacob
  Eisenstein}, {and} \bibinfo{person}{Eric Gilbert}.}
  \bibinfo{year}{2017}\natexlab{}.
\newblock \showarticletitle{You can't stay here: The efficacy of {Reddit}'s
  2015 ban examined through hate speech}. In \bibinfo{booktitle}{\emph{ACM
  CSCW}}.
\newblock


\bibitem[Cresci et~al\mbox{.}(2020)]%
        {cresci2020emergent}
\bibfield{author}{\bibinfo{person}{Stefano Cresci}, \bibinfo{person}{Roberto
  Di~Pietro}, \bibinfo{person}{Marinella Petrocchi}, \bibinfo{person}{Angelo
  Spognardi}, {and} \bibinfo{person}{Maurizio Tesconi}.}
  \bibinfo{year}{2020}\natexlab{}.
\newblock \showarticletitle{Emergent properties, models, and laws of behavioral
  similarities within groups of {Twitter} users}.
\newblock \bibinfo{journal}{\emph{Computer Communications}}
  \bibinfo{volume}{150} (\bibinfo{year}{2020}), \bibinfo{pages}{47--61}.
\newblock


\bibitem[Cresci et~al\mbox{.}(2019)]%
        {cresci2019semantically}
\bibfield{author}{\bibinfo{person}{Stefano Cresci}, \bibinfo{person}{Roberto
  Di~Pietro}, {and} \bibinfo{person}{Maurizio Tesconi}.}
  \bibinfo{year}{2019}\natexlab{}.
\newblock \showarticletitle{{Semantically-aware statistical metrics via
  weighting kernels}}. In \bibinfo{booktitle}{\emph{IEEE DSAA}}.
\newblock


\bibitem[Cresci et~al\mbox{.}(2022)]%
        {cresci2022personalized}
\bibfield{author}{\bibinfo{person}{Stefano Cresci}, \bibinfo{person}{{Trujillo
  A.}}, {and} \bibinfo{person}{Tiziano Fagni}.}
  \bibinfo{year}{2022}\natexlab{}.
\newblock \showarticletitle{{Personalized interventions for online
  moderation}}. In \bibinfo{booktitle}{\emph{ACM HT}}.
\newblock


\bibitem[De~Francisci~Morales et~al\mbox{.}(2021)]%
        {de2021no}
\bibfield{author}{\bibinfo{person}{Gianmarco De~Francisci~Morales},
  \bibinfo{person}{Corrado Monti}, {and} \bibinfo{person}{Michele Starnini}.}
  \bibinfo{year}{2021}\natexlab{}.
\newblock \showarticletitle{No echo in the chambers of political interactions
  on {Reddit}}.
\newblock \bibinfo{journal}{\emph{Scientific reports}} \bibinfo{volume}{11},
  \bibinfo{number}{1} (\bibinfo{year}{2021}).
\newblock


\bibitem[Flores-Saviaga et~al\mbox{.}(2018)]%
        {flores2018mobilizing}
\bibfield{author}{\bibinfo{person}{Claudia Flores-Saviaga},
  \bibinfo{person}{Brian Keegan}, {and} \bibinfo{person}{Saiph Savage}.}
  \bibinfo{year}{2018}\natexlab{}.
\newblock \showarticletitle{Mobilizing the {Trump} train: Understanding
  collective action in a political trolling community}. In
  \bibinfo{booktitle}{\emph{AAAI ICWSM}}.
\newblock


\bibitem[Foucault~Welles(2014)]%
        {welles2014minorities}
\bibfield{author}{\bibinfo{person}{Brooke Foucault~Welles}.}
  \bibinfo{year}{2014}\natexlab{}.
\newblock \showarticletitle{On minorities and outliers: The case for making
  {Big Data} small}.
\newblock \bibinfo{journal}{\emph{Big Data \& Society}} \bibinfo{volume}{1},
  \bibinfo{number}{1} (\bibinfo{year}{2014}).
\newblock


\bibitem[Gelman et~al\mbox{.}(2023)]%
        {gelman2023causal}
\bibfield{author}{\bibinfo{person}{Andrew Gelman}, \bibinfo{person}{Jessica
  Hullman}, {and} \bibinfo{person}{Lauren Kennedy}.}
  \bibinfo{year}{2023}\natexlab{}.
\newblock \bibinfo{booktitle}{\emph{Causal quartets: Different ways to attain
  the same average treatment effect}}.
\newblock \bibinfo{type}{{T}echnical {R}eport}. \bibinfo{institution}{Columbia
  University}.
\newblock
\newblock
\shownote{\url{https://statmodeling.stat.columbia.edu/2023/02/24/causal-quartets-different-ways-to-attain-the-same-average-treatment-effect/}}.


\bibitem[Gerrard(2018)]%
        {gerrard2018beyond}
\bibfield{author}{\bibinfo{person}{Ysabel Gerrard}.}
  \bibinfo{year}{2018}\natexlab{}.
\newblock \showarticletitle{Beyond the hashtag: Circumventing content
  moderation on social media}.
\newblock \bibinfo{journal}{\emph{New Media \& Society}} \bibinfo{volume}{20},
  \bibinfo{number}{12} (\bibinfo{year}{2018}).
\newblock


\bibitem[Gillespie(2018)]%
        {gillespie2018custodians}
\bibfield{author}{\bibinfo{person}{Tarleton Gillespie}.}
  \bibinfo{year}{2018}\natexlab{}.
\newblock \bibinfo{booktitle}{\emph{Custodians of the {Internet}: Platforms,
  Content Moderation, and the Hidden Decisions That Shape Social Media}}.
\newblock \bibinfo{publisher}{Yale University Press}.
\newblock


\bibitem[Hill(1973)]%
        {hill1973diversity}
\bibfield{author}{\bibinfo{person}{Mark~O Hill}.}
  \bibinfo{year}{1973}\natexlab{}.
\newblock \showarticletitle{Diversity and evenness: A unifying notation and its
  consequences}.
\newblock \bibinfo{journal}{\emph{Ecology}} \bibinfo{volume}{54},
  \bibinfo{number}{2} (\bibinfo{year}{1973}).
\newblock


\bibitem[Horta~Ribeiro et~al\mbox{.}(2021)]%
        {horta2021platform}
\bibfield{author}{\bibinfo{person}{Manoel Horta~Ribeiro},
  \bibinfo{person}{Shagun Jhaver}, \bibinfo{person}{Savvas Zannettou},
  \bibinfo{person}{Jeremy Blackburn}, \bibinfo{person}{Gianluca Stringhini},
  \bibinfo{person}{Emiliano De~Cristofaro}, {and} \bibinfo{person}{Robert
  West}.} \bibinfo{year}{2021}\natexlab{}.
\newblock \showarticletitle{Do platform migrations compromise content
  moderation? Evidence from r/The\_Donald and r/Incels}. In
  \bibinfo{booktitle}{\emph{ACM CSCW}}.
\newblock


\bibitem[Jhaver et~al\mbox{.}(2021)]%
        {jhaver2021evaluating}
\bibfield{author}{\bibinfo{person}{Shagun Jhaver}, \bibinfo{person}{Christian
  Boylston}, \bibinfo{person}{Diyi Yang}, {and} \bibinfo{person}{Amy
  Bruckman}.} \bibinfo{year}{2021}\natexlab{}.
\newblock \showarticletitle{Evaluating the effectiveness of deplatforming as a
  moderation strategy on {Twitter}}. In \bibinfo{booktitle}{\emph{ACM CSCW}}.
\newblock


\bibitem[Katsaros et~al\mbox{.}(2022)]%
        {katsaros2022reconsidering}
\bibfield{author}{\bibinfo{person}{Matthew Katsaros}, \bibinfo{person}{Kathy
  Yang}, {and} \bibinfo{person}{Lauren Fratamico}.}
  \bibinfo{year}{2022}\natexlab{}.
\newblock \showarticletitle{Reconsidering tweets: Intervening during tweet
  creation decreases offensive content}. In \bibinfo{booktitle}{\emph{AAAI
  ICWSM}}.
\newblock


\bibitem[Krishnan et~al\mbox{.}(2021)]%
        {krishnan2021research}
\bibfield{author}{\bibinfo{person}{Nandita Krishnan}, \bibinfo{person}{Jiayan
  Gu}, \bibinfo{person}{Rebekah Tromble}, {and} \bibinfo{person}{Lorien~C
  Abroms}.} \bibinfo{year}{2021}\natexlab{}.
\newblock \showarticletitle{Research note: Examining how various social media
  platforms have responded to {COVID-19} misinformation}.
\newblock \bibinfo{journal}{\emph{HKS Misinformation Review}}
  \bibinfo{volume}{2}, \bibinfo{number}{6} (\bibinfo{year}{2021}).
\newblock


\bibitem[Massachs et~al\mbox{.}(2020)]%
        {massachs2020roots}
\bibfield{author}{\bibinfo{person}{Joan Massachs}, \bibinfo{person}{Corrado
  Monti}, \bibinfo{person}{Gianmarco De~Francisci Morales}, {and}
  \bibinfo{person}{Francesco Bonchi}.} \bibinfo{year}{2020}\natexlab{}.
\newblock \showarticletitle{Roots of {Trumpism}: Homophily and social feedback
  in {Donald Trump} support on {Reddit}}. In \bibinfo{booktitle}{\emph{ACM
  WebSci}}.
\newblock


\bibitem[Matakos et~al\mbox{.}(2020)]%
        {matakos2020maximizing}
\bibfield{author}{\bibinfo{person}{Antonis Matakos}, \bibinfo{person}{Cigdem
  Aslay}, \bibinfo{person}{Esther Galbrun}, {and} \bibinfo{person}{Aristides
  Gionis}.} \bibinfo{year}{2020}\natexlab{}.
\newblock \showarticletitle{Maximizing the diversity of exposure in a social
  network}.
\newblock \bibinfo{journal}{\emph{IEEE TKDE}} (\bibinfo{year}{2020}).
\newblock


\bibitem[Molina and Sundar(2022)]%
        {molina2022does}
\bibfield{author}{\bibinfo{person}{Maria~D Molina} {and}
  \bibinfo{person}{S~Shyam Sundar}.} \bibinfo{year}{2022}\natexlab{}.
\newblock \showarticletitle{Does distrust in humans predict greater trust in
  {AI}? Role of individual differences in user responses to content
  moderation}.
\newblock \bibinfo{journal}{\emph{New Media \& Society}}
  (\bibinfo{year}{2022}).
\newblock


\bibitem[Rieder and Skop(2021)]%
        {rieder2021fabrics}
\bibfield{author}{\bibinfo{person}{Bernhard Rieder} {and}
  \bibinfo{person}{Yarden Skop}.} \bibinfo{year}{2021}\natexlab{}.
\newblock \showarticletitle{The fabrics of machine moderation: Studying the
  technical, normative, and organizational structure of {Perspective API}}.
\newblock \bibinfo{journal}{\emph{Big Data \& Society}} \bibinfo{volume}{8},
  \bibinfo{number}{2} (\bibinfo{year}{2021}).
\newblock


\bibitem[Robertson(2022)]%
        {robertson2022uncommon}
\bibfield{author}{\bibinfo{person}{Ronald Robertson}.}
  \bibinfo{year}{2022}\natexlab{}.
\newblock \showarticletitle{Uncommon yet consequential online harms}.
\newblock \bibinfo{journal}{\emph{Journal of Online Trust and Safety}}
  \bibinfo{volume}{1}, \bibinfo{number}{3} (\bibinfo{year}{2022}).
\newblock


\bibitem[Saleem and Ruths(2018)]%
        {saleem2018aftermath}
\bibfield{author}{\bibinfo{person}{Haji~Mohammad Saleem} {and}
  \bibinfo{person}{Derek Ruths}.} \bibinfo{year}{2018}\natexlab{}.
\newblock \bibinfo{title}{The aftermath of disbanding an online hateful
  community}.
\newblock
\newblock
\showeprint[arxiv]{1804.07354}


\bibitem[Shen and Ros{\'e}(2022)]%
        {shen2022tale}
\bibfield{author}{\bibinfo{person}{Qinlan Shen} {and}
  \bibinfo{person}{Carolyn~P Ros{\'e}}.} \bibinfo{year}{2022}\natexlab{}.
\newblock \showarticletitle{A tale of two subreddits: Measuring the impacts of
  quarantines on political engagement on {Reddit}}. In
  \bibinfo{booktitle}{\emph{AAAI ICWSM}}.
\newblock


\bibitem[Singhal et~al\mbox{.}(2022)]%
        {singhal2022sok}
\bibfield{author}{\bibinfo{person}{Mohit Singhal}, \bibinfo{person}{Chen Ling},
  \bibinfo{person}{Nihal Kumarswamy}, \bibinfo{person}{Gianluca Stringhini},
  {and} \bibinfo{person}{Shirin Nilizadeh}.} \bibinfo{year}{2022}\natexlab{}.
\newblock \bibinfo{title}{SoK: Content moderation in social media, from
  guidelines to enforcement, and research to practice}.
\newblock
\newblock
\showeprint[arxiv]{2206.14855}


\bibitem[{Trujillo A.} and Cresci(2022)]%
        {trujillo2022make}
\bibfield{author}{\bibinfo{person}{{Trujillo A.}} {and}
  \bibinfo{person}{Stefano Cresci}.} \bibinfo{year}{2022}\natexlab{}.
\newblock \showarticletitle{{Make Reddit Great Again: Assessing community
  effects of moderation interventions on r/The\_Donald}}. In
  \bibinfo{booktitle}{\emph{ACM CSCW}}.
\newblock


\bibitem[{Trujillo M.} et~al\mbox{.}(2021)]%
        {trujillo2021echo}
\bibfield{author}{\bibinfo{person}{{Trujillo M.}}, \bibinfo{person}{Sam
  Rosenblatt}, \bibinfo{person}{Guillermo de Anda~J{\'a}uregui},
  \bibinfo{person}{Emily Moog}, \bibinfo{person}{Briane Paul~V Samson},
  \bibinfo{person}{Laurent H{\'e}bert-Dufresne}, {and}
  \bibinfo{person}{Allison~M Roth}.} \bibinfo{year}{2021}\natexlab{}.
\newblock \showarticletitle{When the echo chamber shatters: Examining the use
  of community-specific language post-subreddit ban}. In
  \bibinfo{booktitle}{\emph{WOAH}}.
\newblock


\bibitem[Weld et~al\mbox{.}(2022)]%
        {weld2022makes}
\bibfield{author}{\bibinfo{person}{Galen Weld}, \bibinfo{person}{Amy~X Zhang},
  {and} \bibinfo{person}{Tim Althoff}.} \bibinfo{year}{2022}\natexlab{}.
\newblock \showarticletitle{What makes online communities ‘better’?
  Measuring values, consensus, and conflict across thousands of subreddits}. In
  \bibinfo{booktitle}{\emph{AAAI ICWSM}}.
\newblock


\bibitem[Wilkinson et~al\mbox{.}(2016)]%
        {wilkinson2016fair}
\bibfield{author}{\bibinfo{person}{Mark~D Wilkinson}, \bibinfo{person}{Michel
  Dumontier}, \bibinfo{person}{IJsbrand~Jan Aalbersberg},
  \bibinfo{person}{Gabrielle Appleton}, \bibinfo{person}{Myles Axton},
  \bibinfo{person}{Arie Baak}, \bibinfo{person}{Niklas Blomberg},
  \bibinfo{person}{Jan-Willem Boiten}, \bibinfo{person}{Luiz~Bonino da
  Silva~Santos}, \bibinfo{person}{Philip~E Bourne}, {et~al\mbox{.}}}
  \bibinfo{year}{2016}\natexlab{}.
\newblock \showarticletitle{The {FAIR Guiding Principles} for scientific data
  management and stewardship}.
\newblock \bibinfo{journal}{\emph{Scientific Data}} \bibinfo{volume}{3},
  \bibinfo{number}{1} (\bibinfo{year}{2016}), \bibinfo{pages}{1--9}.
\newblock


\bibitem[Williams et~al\mbox{.}(2017)]%
        {williams2017individual}
\bibfield{author}{\bibinfo{person}{Emma~J Williams}, \bibinfo{person}{Amy
  Beardmore}, {and} \bibinfo{person}{Adam~N Joinson}.}
  \bibinfo{year}{2017}\natexlab{}.
\newblock \showarticletitle{Individual differences in susceptibility to online
  influence: A theoretical review}.
\newblock \bibinfo{journal}{\emph{Computers in Human Behavior}}
  \bibinfo{volume}{72} (\bibinfo{year}{2017}).
\newblock


\bibitem[Zannettou(2021)]%
        {zannettou2021won}
\bibfield{author}{\bibinfo{person}{Savvas Zannettou}.}
  \bibinfo{year}{2021}\natexlab{}.
\newblock \showarticletitle{``I Won the Election!'': An empirical analysis of
  soft moderation interventions on {Twitter}}. In
  \bibinfo{booktitle}{\emph{AAAI ICWSM}}.
\newblock


\bibitem[Zannettou et~al\mbox{.}(2020)]%
        {zannettou2020measuring}
\bibfield{author}{\bibinfo{person}{Savvas Zannettou}, \bibinfo{person}{Mai
  ElSherief}, \bibinfo{person}{Elizabeth Belding}, \bibinfo{person}{Shirin
  Nilizadeh}, {and} \bibinfo{person}{Gianluca Stringhini}.}
  \bibinfo{year}{2020}\natexlab{}.
\newblock \showarticletitle{Measuring and characterizing hate speech on news
  websites}. In \bibinfo{booktitle}{\emph{ACM WebSci}}.
\newblock


\end{thebibliography}

\end{document}